\documentclass[12pt]{article} 
\usepackage{psfig}
\def\beq{\begin{equation}} 
\def\eeq{\end{equation}} 
\def\bea{\begin{eqnarray}} 
\def\eea{\end{eqnarray}}
\def\bq{\begin{quote}} 
\def\eq{\end{quote}}
\def\nn{\nonumber}

\def\gappeq{\mathrel{\rlap {\raise.5ex\hbox{$>$}} {\lower.5ex\hbox{$\sim$}}}}

\def\lappeq{\mathrel{\rlap{\raise.5ex\hbox{$<$}} {\lower.5ex\hbox{$\sim$}}}}

\begin{document} 
\pagestyle{empty} 
\begin{flushright} {CERN-TH/2001-0281\\DFPD 01/TH/04} \end{flushright}
\vspace*{5mm}
\begin{center} 
{\bf SU(5) Grand Unification in Extra Dimensions and Proton Decay} \\ 
\vspace*{1cm}  
{\bf Guido Altarelli} \\ 
\vspace{0.3cm} 
Theoretical Physics Division, CERN \\ 
CH - 1211 Geneva 23 \\ 
\vspace{0.3cm} {\bf Ferruccio Feruglio} \\ 
\vspace{0.3cm}  
Universit\`a di Padova\\
and\\
I.N.F.N., Sezione di Padova, Padua, Italy\\
\vspace*{2.0cm}   
{\bf Abstract} \\ 
\end{center}
\vspace*{3mm} 
\noindent
We analyse proton decay in the context of simple supersymmetric SU(5) grand unified models
with an extra compact spatial dimension described by the orbifold $S^1/(Z_2\times Z_2')$.
Gauge and Higgs degrees of freedom live in the bulk, while matter fields can only  
live at the fixed point branes. We present an extended discussion of matter interactions on the brane. 
We show that proton decay is naturally suppressed or even forbidden by suitable implementations of the parity 
symmetries on the brane. The corresponding mechanism does not affect the SU(5) description of fermion
masses also including the neutrino sector, where Majorana mass terms remain allowed. 
\vspace*{1cm}
\noindent  

\begin{flushleft} CERN-TH/2001-028\\DFPD 01/TH/04 \\ February 2001 \end{flushleft} 
\vfill\eject 
\clearpage
\setcounter{page}{1} 
\pagestyle{plain}
\section{Introduction}

The idea that all particle interactions merge into a unified theory at very high energies is so attractive
that this concept has become widely accepted by now. The quantitative success of coupling unification in Supersymmetric (SUSY)
Grand Unified Theories (GUT's) has added much support to this idea. The recent developments on neutrino oscillations,
pointing to lepton number violation at large scales, have further strengthened the general confidence.  However the
actual realization of this idea is not precisely defined. Recently it has been observed that the GUT gauge symmetry could be
actually realized in 5 (or more) space-time dimensions and broken down to the the Standard Model (SM) by compactification
\footnote{Grand unified supersymmetric models in six dimensions, with the grand unified scale related to
the compactification scale were also proposed by Fayet \cite{faye}.}.
In particular a model with N=2 Supersymmetry
(SUSY) and gauge SU(5) in 5 dimensions has been proposed \cite{kawa} 
where the GUT symmetry is broken by compactification on
$S^1/(Z_2\times Z_2')$ down to a N=1 SUSY-extended version of the SM on a 4-dimensional brane. 
In this model many good properties of
GUT's, like coupling unification and charge quantization are maintained while some
unsatisfactory properties of the conventional breaking mechanism, like doublet-triplet splitting, are avoided. In this note
we elaborate further on this class of models. We differ from ref. \cite{kawa} in the form of the interactions on the 4-dimensional brane. As a
consequence we not only avoid the problem of the doublet-triplet splitting but also directly suppress or even forbid proton
decay, since the conventional Higgsino and gauge boson exchange amplitudes are absent, as a consequence of $Z_2\times
Z_2'$ parity assignments on matter fields on the brane.   Most good predictions of SUSY SU(5) are thus maintained without
unnatural fine tunings as needed in the minimal model (for a realistic conventional model, as opposed to a minimal model see
ref. \cite{afm2}). We find that the relations among fermion masses implied by the minimal model, for example
$m_b=m_{\tau}$ at $M_{GUT}$, are preserved in our version of the model, although the Yukawa interactions 
are not fully SU(5) symmetric. The mechanism that forbids proton decay still allows Majorana mass terms for
neutrinos so that the good SU(5) potentiality for the description of neutrino masses and mixing is preserved. 
In the following we first summarise the model of ref. \cite{kawa}, we then introduce our different implementations 
of the boundary conditions and finally we discuss the physical implications of the model and draw our conclusion.

\section{The Model}

Following ref. \cite{kawa} we consider a 5-dimensional space-time factorised into a product of the ordinary 
4-dimensional space-time M$^4$ and of the orbifold $S^1/(Z_2\times Z_2')$, with coordinates $x^{\mu}$, ($\mu=0,1,2,3$) 
and $y=x^5$. The orbifold $S^1/Z_2$ is obtained by dividing the circle $S^1$ of radius $R$ ($1/R\sim M_{GUT}$) with a 
$Z_2$ transformation $y\rightarrow -y$. 
To obtain the orbifold $S^1/(Z_2\times Z_2')$ we divide $S^1/Z_2$ 
by $Z'_2$ which act as $y'\rightarrow -y'$, with $y'=y+ \pi R/2$.
There are two 4-dimensional branes at the fixed points $y=0$, and at $y =\pi R/2$ 
(the brane at $y =- \pi R$ is
identified with that at $y=0$ and those at $y =\pm \pi R/2$ are also identified). For a generic field
$\phi(x^{\mu},y)$ living in the 5-dimensional bulk the $Z_2$ and $Z_2'$ parities $P$ and $P'$ are defined by 
\bea
\phi(x^{\mu},y)&\to \phi(x^{\mu},-y)=P\phi(x^{\mu},y)~~~,\nn\\
\phi(x^{\mu},y')&\to \phi(x^{\mu},-y')=P'\phi(x^{\mu},y')~~~.
\eea
Denoting by $\phi_{\pm \pm}$ the fields with $(P,P')=(\pm,\pm)$ we have the $y$-Fourier expansions:
\bea
  \phi_{++} (x^\mu, y) &=& 
       \sqrt{2 \over {\pi R}} 
      \sum_{n=0}^{\infty} \phi^{(2n)}_{++}(x^\mu) \cos{2ny \over R}~~~,
\label{phi++exp}\nn\\
  \phi_{+-} (x^\mu, y) &=& 
       \sqrt{2 \over {\pi R}} 
      \sum_{n=0}^{\infty} \phi^{(2n+1)}_{+-}(x^\mu) \cos{(2n+1)y \over R}~~~,
\label{phi+-exp}\nn\\
  \phi_{-+} (x^\mu, y) &=& 
       \sqrt{2 \over {\pi R}}
      \sum_{n=0}^{\infty} \phi^{(2n+1)}_{-+}(x^\mu) \sin{(2n+1)y \over R}~~~,
\label{phi-+exp}\nn\\
 \phi_{--} (x^\mu, y) &=& 
       \sqrt{2 \over {\pi R}}
      \sum_{n=0}^{\infty} \phi^{(2n+2)}_{--}(x^\mu) \sin{(2n+2)y \over R}~~~.
\label{phi--exp}
\eea
where $n$ is a non negative integer and the  Fourier component fields $\phi^{(2n)}_{++}$ acquire a mass $2n/R$,
$\phi^{(2n+1)}_{+-}$ and $\phi^{(2n+1)}_{-+}$ take a mass $(2n+1)/R$ and $\phi^{(2n+2)}_{--}$ acquire a mass $(2n+2)/R$,
upon compactification.
Only $\phi_{++}$ have massless components and only $\phi_{++}$ and $\phi_{+-}$ can exist on the 
$y=0$ brane. The fields $\phi_{++}$ and $\phi_{-+}$ are non-vanishing on the $y=\pi R/2$ brane, while $\phi_{--}$
vanishes on both branes. 
\\[0.1cm]
{\begin{center}
\begin{tabular}{|c|c|c|}   
\hline
& & \\                         
$(P,P')$ & field & mass\\ 
& & \\
\hline
& & \\
$(+,+)$ &  $A^a_{\mu}$, $\lambda^{2a}_W$, $H^D_u$, $H^D_d$ & $\frac{2n}{R}$\\
& & \\
\hline
& & \\
$(+,-)$ &  $A^{\hat{a}}_{\mu}$, $\lambda^{2\hat{a}}_W$, $H^T_u$, $H^T_d$ & $\frac{(2n+1)}{R}$ \\
& & \\ 
\hline
& & \\
$(-,+)$ &  $A^{\hat{a}}_5$, $\Sigma^{\hat{a}}$, $\lambda^{1\hat{a}}_W$, $\hat{H}^T_u$, $\hat{H}^T_d$  
& $\frac{(2n+1)}{R}$\\
& & \\
\hline
& & \\
$(-,-)$ &  $A^{a}_5$, $\Sigma^{a}$, $\lambda^{1a}_W$, $\hat{H}^D_u$, $\hat{H}^D_d$ & $\frac{(2n+2)}{R}$ \\
& & \\
\hline
\end{tabular} 
\end{center}}
\vspace{3mm}
Table 1. Parity assignment and masses ($n\ge 0$) of fields in the vector and Higgs supermultiplets.

Our aim is to reproduce most of the good predictions of a minimal (N=1) SUSY SU(5) GUT. For this purpose we start from a 
5-dimensional
theory invariant under N=2 SUSY, gauge SU(5) and $Z_2\times Z_2'$ parities. The parities are assigned in such a way that
compactification reduces N=2 to N=1 SUSY and breaks SU(5) down to the SM  gauge group ${\rm SU(3)} \times {\rm SU(2)} \times
{\rm U(1)}$. Leaving aside for the moment quarks and leptons and their SUSY partners, the 5-dimensional theory contains the
following N=2 SUSY multiplets of fields. First, a vector multiplet (with vector bosons
$A_M$, $M=0,1,2,3,5$, two bispinors $\lambda ^i_W$, $i=1,2$, and a real scalar $\Sigma$, each of them transforming 
as a 24 representation of SU(5). Then there are two hypermultiplets $H^s$ ($s=1,2$) that are equivalent to four sets 
of N=1 chiral multiplets.
Two of them are the ordinary $H_5$ and $H_{\bar 5}$ supermultiplets (which include the scalar Higgs doublets $H^D_u$ and
$H^D_d$ and the corresponding scalar triplets $H^T_u$ and $H^T_d$); the remaining two are denoted by $\hat{H}_5$ and 
$\hat{H}_{\bar 5}$ and will be called mirror. The parity $P$, $P'$ assignments are the same as in ref. \cite{kawa} and are given in table 1.

Here the index $a$ ($\hat{a}$) labels the unbroken (broken) SU(5) generators $T^a$ ($T^{\hat{a}}$), 
$H$ stands for the whole chiral multiplet of given quantum numbers. 
The parity $P$ breaks N=2 SUSY down to N=1 and would allow complete SU(5) massless supermultiplets,
contained in the first two rows of table 1. The additional parity $P'$ respects the surviving N=1 SUSY and breaks
SU(5) down to the standard model gauge group. Note that the derivative $\partial_5$ 
transforms as $(-,-)$. The $(+,+)$ fields, which remain
massless and do not vanish on both branes are the gauge and Higgs multiplets of the low 
energy Minimal SUSY Model (MSSM). The bulk 5-dimensional Lagrangian is exactly as in ref. \cite{kawa} and we do not 
reproduce it here. 

If we expand the bulk Lagrangian around the local minimum $A_5=\Sigma=0$ of the scalar
potential, then the mass terms are only those provided by the derivatives $\partial\phi/\partial y$
of the fields $\phi$. The resulting spectrum is the one shown in table 1.
An Higgs mechanism occurs at each level $n\ne 0$. The vector bosons $A_\mu^{a(0)}$
remains massless and, together with the gauginos $\lambda_W^{2a(0)}$, form a vector multiplet
of N=1. The vector bosons $A_\mu^{a(2n)}~~~(n>0)$ eat the Goldstone bosons $A_5^{a(2n)}$, acquire
a mass $2n/R$ and, together with $\Sigma^{a(2n)}$, $\lambda_W^{1a(2n)}$ and $\lambda_W^{2a(2n)}$,
form a massive vector multiplet of N=1. Likewise, full N=1 massive vector multiplets are
reproduced by $\Sigma^{\hat a(2n+1)}$, $\lambda_W^{1\hat a(2n+1)}$, 
$\lambda_W^{2\hat a(2n+1)}$ and by $A_\mu^{\hat a(2n+1)}$ that acquire a mass $(2n+1)/R$
by eating the Goldstone bosons $A_5^{\hat a(2n+1)}$. 

Note that mass terms for the Higgs hypermultiplets are allowed by N=2 SUSY \cite{sohn},
but they are not compatible with the $P$ parity assignment of the fermion fields:
the Lagrangian would contain both ${\tilde H}_5 \Sigma {\tilde{\hat H}}_{\bar 5}$
and ${\tilde H}_5 {\tilde{\hat H}}_{\bar 5}$ terms, where a tilde denotes the fermion component
of the corresponding chiral multiplet. But the two sets of terms cannot be both allowed given
the negative $P$ parity of
$\Sigma$, which is uniquely determined by the gauge sector alone. Therefore the $P$ parity symmetry
forbids a mass term for any hypermultiplet. As a consequence, the zero modes of the doublets 
$H^D_{u,d}$ cannot receive a bulk mass contribution and are automatically split from
the $H^T_{u,d}$ modes.

\section{Matter Fields}

We now discuss the introduction of quark and lepton matter. In this sector our formulation of the model is different than
in ref. \cite{kawa}. We first observe that it is not possible that the matter fields propagate and interact in the bulk. To
introduce quark and lepton fields in the bulk they should enter in the Lagrangian in a N=2 SUSY, gauge SU(5) and 
$Z_2\times Z_2'$ invariant form. Thus we should define N=2 hypermultiplets $\chi_{\bar 5}$,
$\chi_{10}$ (and possibly $\chi_1$). Each N=2 hypermultiplet $\chi$ contains a N=1 chiral multiplet 
$\psi$, $\phi$ of fermions and scalars and another chiral multiplet $\hat{\psi}$, $\hat{\phi}$. 
The ordinary quark and lepton supermultiplets must remain massless and must be non-vanishing on branes. Thus the parity assignments must be $(+,+)$ for all SU(5) indices of
$\psi_{\bar 5}$, $\phi_{\bar 5}$, $\psi_{10}$, $\phi_{10}$ (and possibly of $\psi_1$, $\phi_1$). 
Thus, contrary to what is the case for the vector and Higgs hypermultiplets, 
for the $\bar 5$ and $10$ chiral supermultiplets 
the whole SU(5) representations have the same parity transformation. Due to this fact, it is straightforward to prove that
it is impossible to construct gauge SU(5) and
$Z_2\times Z_2'$ invariant interactions of matter fields with the gauge and Higgs supermultiplets in the bulk. 
For example, this can be seen by inspecting the term $\psi_{\bar 5} \Sigma\hat{\psi}_5$. 
(In N=2 SUSY all interactions are gauge interactions and this term is part of them.)
If we decompose the pentaplets in their
doublet (D) and triplet (T) components, this term can be symbolically written as:
\beq
\psi^T \Sigma^a  \hat{\psi}^T 
+\psi^D \Sigma^a  \hat{\psi}^D
+\psi^T \Sigma^{\hat a} \hat{\psi}^D
+\psi^D \Sigma^{\hat a} \hat{\psi}^T~~~~~.
\nn
\eeq
Positive $P'$ for both matter fields $\psi^T$ and $\psi^D$ cannot be obtained whatever $P'$ parity we
may assign to the mirror fields $\hat{\psi}^T$ and $\hat{\psi}^D$. Hence the matter fields have necessarily to be 
introduced only on the fixed point branes.

Interaction terms between bulk and brane fields have been discussed in ref. \cite{shar,mira} and explicitly
worked out for SM matter in ref. \cite{delg}. An important point is the symmetry that should be respected
by these interactions. To analyse this in detail, it is instructive to build the theory in two steps, by first
constructing the orbifold $S^1/Z^2$ and by subsequently performing the orbifold identification
required by $Z_2'$. The first $Z_2$ leaves both the SU(5) gauge symmetry and a residual N=1 supersymmetry
unbroken. Matter interactions, located at the fixed points $(0,\pi R)$, are allowed to respect these
symmetries \cite{shar,mira,delg,quir,barb}. In particular we can introduce SU(5) invariant interactions 
between matter and gauge
fields and also SU(5) invariant Yukawa couplings on the branes at $y=(0,\pi R)$. No relation
between the interactions on the two different branes is required at this stage.
In the following we take the point of view that on the $y=(0,\pi R)$ branes matter must transform
according to complete representations of SU(5) and all interactions will have SU(5) symmetric couplings.
When the $y$ integration is performed
the breaking of SU(5) will be generated by imposing $P'$ invariance  
({\em i.e.} terms that are odd under $y=0\leftrightarrow y=\pi R$ disappear), as we now illustrate.
When we identify $y'$ and $-y'$ and we enforce the $Z_2'$ discrete symmetry, we are no longer allowed
to consider the two fixed points $y=(0,\pi R)$ separately. Indeed they are now related
and we expect that interaction terms on one brane are not independent from those on the other brane. 
Consider for instance the gauge couplings induced on the brane at $y=0$ by the Kahler potential:
\beq
K=10^\dagger e^V 10+{\bar 5}^\dagger e^V {\bar 5}~~~~~.
\eeq
Here $10\equiv (Q,U^c,E^c)$ and ${\bar 5}\equiv (L,D^c)$ are N=1 chiral multiplets, 
$V\equiv (A_\mu,\lambda_W^2)$ is an N=1 vector multiplet, with obvious meaning of the notation.
Each chiral multiplet $\Phi_M\equiv (\phi_M,\psi_M)$, $M=(Q,L,U^c,D^c,E^c)$, contains 
a complex scalar $\phi_M$ and a two-component fermion $\psi_M$.
The corresponding fermion interactions, in a compact notation, are given by:
\beq
{\cal L}_{g}={\cal L}^a_{g}+{\cal L}^{\hat a}_{g}~~~,
\eeq
\beq
{\cal L}^a_{g}=\sum_M \overline{\psi_M} \bar\sigma^\mu T^a \psi_M A^a_\mu~~~,
\label{gauge1}
\eeq
\beq
{\cal L}^{\hat a}_{g}=
\overline{\psi_Q} \bar\sigma^\mu T^{\hat a} \psi_{U^c} A^{\hat a}_\mu+
\overline{\psi_Q} \bar\sigma^\mu T^{\hat a} \psi_{E^c} A^{\hat a}_\mu+
\overline{\psi_L} \bar\sigma^\mu T^{\hat a} \psi_{D^c} A^{\hat a}_\mu+{\rm h.c.}~~~~~.
\label{gauge2}
\eeq
where $\bar\sigma_\mu=(1,-\sigma^k)$ and $\sigma^k$ are the Pauli matrices. 
Before the $Z_2'$ orbifolding, the gauge symmetry on the $y=0$ brane 
is the largest possible one,
compatible with the full SU(5) group, in agreement with the fact that the $Z_2$ orbifold leaves SU(5)
invariant. If we now perform a $P'$ parity transformation, the fixed point $y=0$ is mapped in $y=\pi R$
and, to obtain a $Z_2'$ invariant action, we should put on the brane at $y=\pi R$ the same gauge 
interactions of eq.
(\ref{gauge1},\ref{gauge2}). We obtain the effective four-dimensional Lagrangian ${\cal L}^{(4)}_g$:
\beq
{\cal L}^{(4)}_g=\int dy \left[\delta(y)+\delta(-y+\pi R)\right] {\cal L}_{g}(y)~~~.
\label{int}
\eeq
At this point we have different possibilities, depending on the $P'$ parity
of matter fields, that specifies how a matter field on the brane at $y=0$
is related to the same field on the brane at $y=\pi R$. For generic 
$P'$ parity assignments SU(5) is broken as a result of the integration in eq. (\ref{int}).
In particular this is the case if all matter fields are assumed to be invariant under $Z_2'$. 
Then the terms in eq. (\ref{gauge1})
are invariant under $y'\to -y'$, that is $y\to -y+\pi R$, whereas those in eq. (\ref{gauge2}) 
change sign. After
integrating over $y$ the terms of eq. (\ref{gauge2}) cancel and we are left with gauge interactions
of matter with the vector bosons of ${\rm SU(3)}\times {\rm SU(2)}\times$ U(1).
The SU(5) invariance has been lost. Only the invariance under the subgroup left unbroken by $Z_2'$
survives.

Alternatively we can try to preserve SU(5) by assigning suitable non-trivial  
parities under $Z_2'$. In other words we allow for a change of sign between a matter field on the brane at $y=0$
and the same field on the brane at $y=\pi R$. In this way
we can preserve SU(5) invariant gauge interactions on the branes at $y=(0,\pi R)$.
Indeed, the $P'$ parity transformation does not destroy the SU(5) algebra. 
It effectively induces a redefinition of the generators according to 
$(T^a,T^{\hat a})\to ({T^a}',{T^{\hat a}}')=(T^a,-T^{\hat a})$.
The old generators $T$ and the new ones $T'$ satisfy the same algebra. 
Therefore, if we assign opposite $P'$ parities to $Q$ and $U^c$, $E^c$ and similarly
for $L$ and $D^c$, the gauge interactions of eq. (\ref{gauge1},\ref{gauge2}) are $Z_2'$ invariant
and, after the integration over $y$, we obtain full SU(5) four-dimensional gauge interactions
for matter fields. This possibility can be realized for the gauge couplings in eqs. (\ref{gauge1},\ref{gauge2}),
but we shall see that it does not lead to a viable alternative for Yukawa couplings.

This discussion shows that the theory is not completely defined unless we specify the
transformation properties of matter fields at $y=(0,\pi R)$ under $Z_2'$. 
Having matter fields invariant under $Z_2'$ represents just one of the possible choices, 
not necessarily the most interesting or the most realistic one 
\footnote{Likewise, when considering matter located at $y=\pm \pi R/2$, we must define the $Z_2$ parities
of the corresponding fields.}.
It is also clear that in general SU(5) invariance of brane interactions is not guaranteed.
It depends on how the matter fields transform under $Z_2'$. For generic $P'$ parities
of matter multiplets only the invariance under the group unbroken by the compactification, 
${\rm SU(3)}\times {\rm SU(2)}\times$ U(1), is recovered.
Notice that in both of the cases discussed above the $Z_2\times Z_2'$ parity,
that plays an essential role in the orbifold construction, is conserved.

We now analyse some of the possible parity assignments for the matter fields and 
discuss the related physics, focusing on the features of proton decay.
We will show that along these lines we can construct a model where not only the doublet-triplet
splitting problem is solved but also proton decay is drastically suppressed or even forbidden in a natural way.

\section{Matter interactions on the branes}

We now introduce Yukawa couplings on the branes. We start from the interactions induced on the brane 
at $y$ by the SU(5) invariant superpotential
\footnote{The Lagrangian terms are derived from the superpotential $w$ with the procedure described in \cite{shar,mira,delg},
entailing the inclusion of appropriate  couplings between brane fields and the $\partial_5$ derivative of bulk
fields.}:
\beq
w(y)=y_u~ 10~ 10~ H_5 + y_d~ 10~ {\bar 5}~ H_{\bar 5}+y_R~ 10~ {\bar 5}~ {\bar 5}+...
\eeq
where $y_{u,d,R}$ are coupling constants and dots denote terms depending on mirror Higgs multiplets. The usual decomposition holds:
\beq
10~ 10~ H_5= Q U^c H_u^D + \frac{1}{2} Q Q H_u^T + U^c E^c H_u^T~~~,\nn
\eeq
\beq
10~ {\bar 5}~ H_{\bar 5}= Q D^c H_d^D + L E^c H_d^D + Q L H_d^T + U^c D^c H_d^T~~~,\nn
\eeq
\beq
10~ {\bar 5}~ {\bar 5}= Q L D^c + E^c L L + U^c D^c D^c~~~.\nn
\eeq
Yukawa couplings invariant under $Z_2'$ are obtained by requiring the same superpotential $w$ on the branes
at $y=(0,\pi R)$:
\beq
w^{(4)}=\int dy \left[\delta(y)+\delta(-y+\pi R)\right] w(y)~~~.
\label{intyuk}
\eeq
We first discuss the possibility of preserving full SU(5) symmetry on the branes. 
It is easy to see that this does not lead to a realistic
model. As we have seen before, the only assignments of $P'$ parities to matter compatible with 
SU(5) symmetry of the gauge couplings are 
$(Q,L,U^c,D^c,E^c)\sim \pm(+,+,-,-,-)$ and $(Q,L,U^c,D^c,E^c)\sim \pm(+,-,-,+,-)$.
The first possibility however leads to $P'$-odd Yukawa couplings that cancel after integrating over $y$ the
two contributions in eq. (\ref{intyuk}).  In the second case, only the term $10~ {\bar 5}~ H_{\bar 5}$
is $P'$-even and survives in the four-dimensional Lagrangian. The other two interactions, 
$10~ 10~ H_5$ and $10~ {\bar 5}~ {\bar 5}$ are $P'$-odd and eventually cancel.
Thus, masses for up type quarks cannot be generated in this model.
As a consequence, we must abandon the possibility of SU(5)-invariant brane interactions and look for
a more realistic choice of $P'$ parity for matter.

As we saw in the case of gauge interactions, for generic parity assignments of matter fields, 
the four-dimensional interactions on the fixed point branes at $y=(0,\pi R)$ only respect the residual 
symmetries after compactification, that is N=1 SUSY, gauge
${\rm SU(3)} \times {\rm SU(2)} \times {\rm U(1)}$ and $Z_2\times Z_2'$ parity.  
In general on the branes also interactions involving $\partial_5$ appear. At first we omit
these derivative terms and we will comment on them later in the following. We also do not include all 
higher dimensional operators
at this stage. We start by observing that the $(P,P')$ parities of the matter supermultiplets
$Q$, $U^c$, $D^c$, $L$ and $E^c$ should allow at least the superpotential terms that
provide masses to the observed fermions (neglecting neutrino masses for the time being):
\beq 
w_{mass}=Q U^c H_u^D + Q D^c H_d^D + L E^c H_d^D~~~.
\nn
\eeq 
Recalling that $H_u^D$ and $H_d^D$ have parities $(+,+)$ it is clear that $Q$, $U^c$ and $D^c$ should have 
equal parities
and similarly for $L$ and
$E^c$. The points $y=(0,\pi R)$ are kept fixed by a $Z_2$ transformation. It is then natural to choose 
positive $P$ parity for all matter fields located on these branes.
With this choice of parities that guarantee non-vanishing fermion masses, 
we now examine the features of proton decay. 
It is well known that in the minimal SUSY SU(5) model the exchange of coloured
Higgsinos can induce proton decay  at a rate which is difficult to reconcile with present experimental limits \cite{plim}. The
superpotential terms describing the corresponding amplitudes are:
\beq 
H_u^T H_d^T~~~,
\label{mix}
\nn
\eeq
together with
\bea
Q Q H_u^T~~&,~~~~~& U^c D^c H_d^T~~,\nn\\
Q L H_d^T~~&,~~~~~& U^c E^c H_u^T~~.
\label{yuk}
\eea
In minimal SUSY SU(5) the B/L-violating dimension 5 operators
\footnote{As usual we denote by the suffix $F$ ($D$)
the integration over $d^2\theta$ ($d^2\theta d^2{\bar\theta}$) in superspace.}
$[QQQL]_F$ and $[U^c U^c D^c E^c]_F$
arise from the trilinear terms of eq. (\ref{yuk}) in the low-energy limit,
after integration over the heavy supermultiplets $H^T_{u,d}$ whose mixing is 
provided by (\ref{mix}).
In the model under discussion the bilinear brane couplings $H_u^T H_d^T$
and ${\hat H}_u^T {\hat H}_d^T$ are allowed by the parity symmetry.
On the contrary, the couplings in the first line of eq. (\ref{yuk}) 
have always parity $(+,-)$ and drop from the final four-dimensional action. A similar 
consideration holds for the analogous couplings obtained with the 
replacements $H_{u,d}^T\to {\hat H}_{u,d}^T$.
Note that $\partial_5$ derivatives being $(-,-)$ cannot rescue 
these couplings.  Terms in $\partial _5$ of $(-,-)$ fields can appear on the brane but they do not have important consequences 
because the lightest $(-,-)$ states are heavy. Therefore, while the $H^T_{u,d}$,  
${\hat H}_{u,d}^T$ masses are of order $M_{GUT}$ as in the 
conventional model, nevertheless their allowed interactions with matter 
are not sufficient to give rise to the dangerous dimension five $[QQQL]_F$ 
and $[U^c U^c D^c E^c]_F$ operators. 

While the dimension 5 operators
$[QQQL]_F$, 
$[U^c U^c D^c E^c]_F$ cannot arise from tree-level bulk field exchange, they may be already present on the brane as
non-renormalizable interactions induced, for instance, by new physics at the Planck scale. On dimensional grounds, these
interactions would lead to unacceptably fast proton decay \cite{ehnt}. 
It is interesting to note that these operators, can
be directly forbidden by a suitable parity assignment, for instance by taking, at
$y=(0, \pi R)$:
\beq
(Q,U^c,D^c)~\sim~(+,+)~~;~~~~~~(L,E^c)~\sim~(+,-)~~~.\label{pp}
\eeq
This is an assignment that we consider in the following as a particularly interesting example. With this choice proton decay is
actually forbidden.

We see from eq. (\ref{gauge2}) that
a non-vanishing coupling $\overline{\psi_Q} \bar\sigma^\mu T^{\hat a} \psi_{U^c} A^{\hat a}_\mu$ is a 
necessary condition for proton decay to occur through gauge vector boson exchange.
However the requirement of non-vanishing Yukawa couplings forces $Q {U^c}^\dagger$ to be $(+,+)$, while
$A_\mu^{\hat a}$ is $(+,-)$. Therefore the coupling 
$\overline{\psi_Q} \bar\sigma^\mu T^{\hat a} \psi_{U^c} A^{\hat a}_\mu$ drops from the total four-dimensional
action. This forbids the occurrence of dimension six operators 
$[QQ{U^c}^\dagger {E^c}^\dagger]_D$ and 
$[{U^c}^\dagger {D^c}^\dagger QL]_D$, for any parity choice of matter fields that guarantees the
invariance of the superpotential $w_{mass}$. In particular, such operators are also
directly forbidden by our example of parity assignment given in eq. (\ref{pp}). 

To avoid fast proton decay we should also dispose of at least some of the terms 
\beq
Q D^c L~~,~~~~~~ L E^c L~~,~~~~~~ U^c D^c D^c~~~~.
\eeq
For this purpose, in the conventional approach, one imposes 
the usual R-parity, as an additional symmetry. In our case we immediately see that the first two vertices are
$P'$-odd according to
the parity assignment in eq. (\ref{pp}). The remaining one cannot by itself lead to proton decay.  

From eq. (\ref{intyuk}) we see that the parity assignment in eq. (\ref{pp}) gives rise to the following 
four-dimensional superpotential:
\bea
w^{(4)}&=& 2 \int dy \delta(y) \left[ y_d~(Q D^c H_d^D + L E^c H_d^D + Q L H_d^T)+\right.\nn\\
& & \left. ~~~~~~~~~~~~~~~y_u~(Q U^c H_u^D + U^c E^c H_u^T)+y_R~ U^c D^c D^c\right]~~~.
\eea 
In particular, specific properties of SU(5) Yukawa couplings
are maintained, like the minimal SU(5) relation $m_d=m^T_e$. Considering now the neutrino mass sector, for each family, we introduce an SU(5)-singlet right-handed neutrino field 
$\nu^c$ on
the branes $y=(0,\pi R)$ and attribute to it the same intrinsic parities as for $L$ and $E^c$, 
for example $(+,-)$ in the case of
eq. (\ref{pp}). Then the Yukawa interaction term $L\nu^c H^D_u +D^c \nu^c H^T_u$, (that leads to Dirac mass terms after electroweak
symmetry breaking) and the $\nu^c$ Majorana mass terms $M_R  \nu^c \nu^c$ are both allowed by parities. The mass $M_R$
is naturally of order $M_{GUT}$ as the $\nu^c$ Majorana mass is compatible with the low energy symmetry. Thus the usual see-saw
mechanism remains viable in this model. Similarly the higher dimensional operator 
$(\lambda/M_L) L H_d^D L H_d^D$, which leads
to light neutrino Majorana mass terms upon electroweak symmetry breaking, is also allowed, with a mass $M_L$ naturally of order
$M_{GUT}-M_{Planck}$. Thus we see that the mechanisms that suppress of even forbid  proton decay leave the good qualitative
features of the neutrino mass sector unaltered. Moreover, the relation between left-handed mixings for charged leptons and right-handed mixings for down quarks, which plays an important role in reproducing large
mixings for neutrinos \cite{lmix}, is  preserved as a consequence of $m_d=m^T_e$.

Summarising, we have the following interesting properties of the interactions on the branes at $y=(0,\pi R)$
\footnote{Matter fields and interactions can also be confined
on the branes located at $y=(\pm \pi R/2)$. Those branes are only relevant after the
$Z_2'$ compactification and in this case the assumption of starting
from SU(5) invariant interactions 
is unjustified. Nevertheless the tree-level suppression of proton decay is still valid. 
The fields $A^{\hat a}_\mu$ and $H_{u,d}^T$ all vanish on these branes and any interaction involving
these fields is automatically absent. Also the interactions involving ${\hat H}_{u,d}^T$
are not dangerous for proton decay. Indeed these fields are $P$-odd and the couplings 
$Q Q {\hat H}_u^T$, $U^c D^c {\hat H}_d^T$ again cancel if non-zero fermion masses can be generated.}. 
First of all, any parity assignment of matter fields that allows non-vanishing fermion masses
forbids tree-level Higgsino or gauge boson exchange amplitudes for proton decay.
Proton decay can be actually forbidden by a suitable set of parity assignments, as those given in (\ref{pp}). 
The qualitative good features induced by lepton number non-conservation in the neutrino mass
sector are preserved. The doublet-triplet splitting problem is solved, as observed in ref. \cite{kawa}. The gauge couplings on the brane of the matter fields are those induced by
${\rm SU(3)}\times {\rm SU(2)}\times {\rm U(1)}$ gauge invariance. The gauge coupling unification and charge
quantisation constraints are valid on the brane because they are guaranteed by the fact that  gauge 
multiplets interact in the bulk. 
The Yukawa interactions that provide fermion masses after electroweak symmetry breaking 
satisfy the minimal SU(5) symmetry relations. Thus mass
relations like $m_b=m_{\tau}$ at $M_{GUT}$ are preserved in this approach.
A $\mu$ term can technically be implemented on the brane as it is compatible with all the unbroken symmetries. The
smallness of the $\mu$ term is a naturalness problem that has no solution in this context.

\section{Conclusion}

The beauty and elegance of grand unification clash with their specific realization in the 
context of conventional models in four space-time dimensions. Realistic models with natural doublet-triplet splitting
and proton decay compatible with the present experimental limits are rather complicated \cite{afm2}.
We find very attractive that in a simple SU(5) model, with a compact fifth dimension,
gauge symmetry breaking and doublet-triplet splitting are accomplished
without the need of a baroque Higgs sector \cite{kawa}. The gauge symmetry is broken
by the action of a $Z_2'$ discrete symmetry and the inverse radius of the
fifth dimension provides the grand unified mass scale. In this note we have analysed
in detail matter interactions in this class of models. We have
shown that matter fields should necessarily live on the branes, where Yukawa
couplings are localised. Next we have shown that, when considering matter interactions located
at the points invariant under $Z_2$ $(Z_2')$, the theory is fully defined only when
the transformation properties of matter under $Z_2'$ $(Z_2)$ are specified.
Four-dimensional matter couplings invariant under the full SU(5) group are in principle allowed, 
but they lead to vanishing up-type quark masses
and are thus ruled out. As a consequence, the interactions only respects
${\rm SU(3)}\times {\rm SU(2)}\times {\rm U(1)}$.
Proton decay cannot proceed via tree-level Higgsino or gauge boson exchange,
for any matter parity compatible with non-vanishing fermion masses. 
Appropriate parity assignments
to matter fields can also forbid four and five dimensional B/L violating
operators, thus playing the role of $R$ parity symmetry. 
Fermion mass relations of minimal SU(5), like $m_b=m_\tau$, are preserved.
Both Dirac and Majorana neutrino masses can be included and see-saw
masses for light neutrinos can be reproduced. A large mixing for neutrinos
can still arise from a large mixing between right-handed quarks, thanks to
the relation $m_e=m_d^T$.
Gauge coupling unification and charge quantization are guaranteed by the SU(5) bulk symmetry.

In the models discussed here
the absence of tree-level amplitudes leading to proton decay appears deeply entangled
with the orbifold construction and it can provide an explanation
to the present negative experimental results. Proton decay could still
proceed through higher dimensional non-renormalizable operators whose
suppression is not a model independent feature, but 
a particular parity assignment of matter fields can forbid proton decay at all. 
The idea of forbidding proton decay by a suitable discrete symmetry is not new \cite{wein},
but the physical origin of the relevant parity is particularly clear in the present context.

Of course there is a long way to the
construction of a realistic model. In this note we have left aside some crucial issues 
like the breaking of the residual N=1 SUSY, realistic fermion masses (including neutrinos) and 
threshold corrections to gauge coupling unification. However we find very encouraging
that at the simplest level the main problems that plague minimal versions
of grand unified theories, like the doublet-triplet splitting and 
fast proton decay, can be overcome rather easily.

\noindent

{\bf Acknowledgements}

\noindent
We would like to thank Ignatios Antoniadis, Andrea Brignole, Mariano Quiros, Claudio Scrucca and Fabio
Zwirner for useful discussions.

\end{document}